\documentclass[letter]{aa} 
\usepackage{graphicx}
\usepackage{txfonts}
\usepackage{natbib}
\bibpunct{(}{)}{;}{a}{}{,} 
\def\deg{\ifmmode^\circ\else$^\circ$\fi}

\def\lsun{L$_{\odot}$}
\def\brg{Br$\gamma$}

\newcommand{\mic}{\,$\mu$m}

\def\n2h+{N$_2$H$^+$}
\def\c18o{C$^{18}$O}

\def\deg{\ifmmode^\circ\else$^\circ$\fi}

\begin{document}
   \title{Probing the close environment of massive young stars with spectro-astrometry}

   \subtitle{}

   \author{J. M. C. Grave\inst{1,2}
          \and M. S. N. Kumar\inst{1}
		}

   \offprints{J. M. C. Grave; email:jgrave@astro.up.pt}

   \institute{Centro de Astrof\'{i}sica da Universidade do Porto, 
              Rua das Estrelas, 4150-762 Porto, Portugal
	  \and Departamento de Matem\'{a}tica Aplicada da Faculdade de Ci\^{e}ncias da Universidade do Porto, Portugal\\
	  \email{jgrave@astro.up.pt; nanda@astro.up.pt}}

   \date{}

\abstract
{} {We test the technique of spectro-astrometry as a potential method
to investigate the close environment of massive young stars.} 
{Archival VLT near infrared K band spectra (R=8900) of three massive
young stellar objects and one Wolf-Rayet star are examined for
spectro-astrometric signatures. The young stellar objects display
emission lines such as \brg, CO 2-0 and CO 3-1 that are characteristic
of ionised regions and molecular disks respectively. Two of the sample
sources also display emission lines such as NIII and MgII that are
characteristic of high temperatures.}  { Most of the emission lines
show spectro-astrometric signal at various levels resulting in
different positional displacements. The shapes and magnitudes of the
positional displacements imply the presence of large disk/envelopes in
emission and expanding shells of ionised gas. The results
obtained for the source 18006-2422nr766 in particular provide larger
estimates ($>$ 300AU) on CO emitting regions indicating that in
MYSOs CO may arise from inner regions of extended dense envelopes as
well.}  { The overall results from this study demonstrate the utility
of spectro-astrometry as a potential method to constrain the sizes of
various physical entities such as disks/envelopes, UCHII regions
and/or ionised shells in the close environment of a massive young
star.}


   \maketitle
%

\section{Introduction}

The technique of spectro-astrometry (SA) has been explored in the past
few years due to its potential in addressing important astrophysical
problems. It has been successfully applied to the detection of
pre-main sequence binaries {\citep{Bail98b,Tak03,Bain06}}, the study
of young stellar object (YSO) jets
\citep{Garcia99,Tak01,Tak02,Tak03,Whel04,Dav03} and the disc structure
{\citep{Tak01,Tak03}}. The main advantage of SA technique is that it
allows us to trace the spatial structure of an unresolved object at
scales well below the seeing or diffraction limit of the
telescope. This is achieved by measuring the centre of the PSF as a
function of wavelength (known as a position spectrum) and measuring
the displacements of emission and/or absorption feature centroids with
respect to the centroid of the continuum. The presence of any
displacements imply a different origin of the line emission/absorption
features with respect to the continuum source. In a theoretical
scenario, the accuracy of these measurements will only depend on
photon statistics, i.e., the measured error will be proportional to
the seeing length of the image and inversely proportional to the
square root of the detected flux
\citep{Bail98a}. For normal sky conditions, a well exposed image can
give an accuracy of a few milli-arcseconds. Detailed studies of the
closest young stars using this technique have resulted in new insights
of the close environments of low-mass YSOs.

In the studies of star formation, massive young stars remain poorly
understood. The close environment of a massive young star is much more
complex than that of a low mass star due to the additional presence of
ultra-compact HII (UCHII) regions and ionised winds along with disks
and outflows. The relatively larger distances at which they are
generally situated have made observations of these complexities very
difficult. In this paper we will try to exploit the potential of SA in
unveiling the complex structure of the close environment of young
massive stars. Our method involves measuring the spatial asymmetry of
various emission lines representative of different physical entities
such as disk, UCHII region and jets/winds using the SA technique. For
this purpose we have utilised the VLT archival near-infrared (NIR)
spectra available as a result of a NIR spectroscopic survey of massive
young stellar objects (MYSOs) by \citet{Bik06}.

\section{Data selection and analysis}

The \citet{Bik06} survey consists of observations with the ISAAC
camera mounted on VLT's UT1, using a long slit (120$\arcsec$) with a
width of 0.3$\arcsec$ resulting in a spectral resolution of R=8900. We
obtained these data from the ESO Science Archive Facility. Although
several tens of source spectra are available from this survey, the
intense nebular emission from the HII regions surrounding young
massive stars and source multiplicity limits the usable sample for the
purpose of SA. In excluding such effects, we found eight relatively
clean single sources with low contamination from the nebular lines
that were suitable for SA analysis. Of these eight sources, four
sources, namely 08576-4334nr292, 18449-0158nr335, 18006-2422nr766 and
16164-5046nr3636 reveal SA signals that are discussed
here. 18449-0158nr335 is a Wolf-Rayet star whereas the other three
sources are MYSOs. K band spectra in the range 2.069\mic\--2.189\mic\
were available for all the sources. For all but 18449-0158nr335,
spectra centred at 2.234 \mic\ covering the first CO overtone bands
were also available.

The raw data obtained from the archive were reduced using standard
spectroscopic reduction techniques. Fringing effects were carefully
removed and the SA analysis was carried out on each exposed
frame. This is necessary because errors in combining the individual
exposures in a standard way can actually spoil the SA signal. Position
spectra were extracted from individual exposures of each source and
then combined together in order to improve the S/N of the SA
signal. To cancel instrumental effects in the position spectra and
obtain a reliable SA signal, ideally one has to obtain spectra in
anti-parallel position angles on the same source
\citep{Bail98b}. This is specially relevant to remove artifacts on the
position spectrum of the emission/absorption lines. However, even when
spectra were limited to one or two position angles, SA analysis has
been successfully carried out by fitting a polynomial function to the
continuum position and subtracting it out from the raw position
spectrum \citep{Dav03,Whel04}. This allows to correct from curvature
and tilt of the spectral image. In our trial case we follow this
method in the following way: the raw position spectrum was divided by
the weighting factor of the continuum which is defined as
\begin{equation}
   I_{\lambda(line)}/[I_{\lambda(line)}+I_{\lambda(cont)}]=[1+I_{\lambda(cont)}/I_{\lambda(line)}]^{-1},   
\end{equation} 
where $I_{\lambda(line)}$ is the intensity of the line and $I_{\lambda(cont)}$ is the intensity of the continuum \citep{Tak01}. The final position $X_{\lambda}$ is obtained by
\begin{equation}
X_{\lambda}=(X_{\lambda(raw)}-X_{\lambda(cont)})\cdot(1+I_{\lambda(cont)}/I_{\lambda(line)}) 
\end{equation}
where $X_{\lambda(raw)}$ is the raw position and $X_{\lambda(cont)}$
is the fitted continuum position. The accuracy of the final position
spectrum is obtained by computing the error on Equation.~2. Only those
signals that were above 10\% of the continuum level are treated as
lines for computing the position spectra. Any signal below this level
is simply treated as continuum.

   \begin{figure}
   \centering
   \includegraphics[width=9cm]{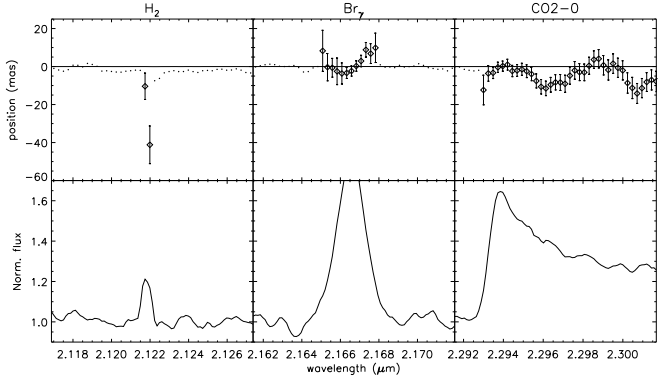}
   
   \caption{Position spectra (top) and normalised intensity spectra
   (bottom) for the lines H$_{2}$, Br$\gamma$ and CO 2-0 bandhead of
   the source 08576-4334nr292. In the top panel, diamonds correspond
   to the position spectra and dots represent the raw continuum
   positions. The solid line shows the fitted continuum position.}
\end{figure}

  \begin{figure} \centering \includegraphics[width=9cm]{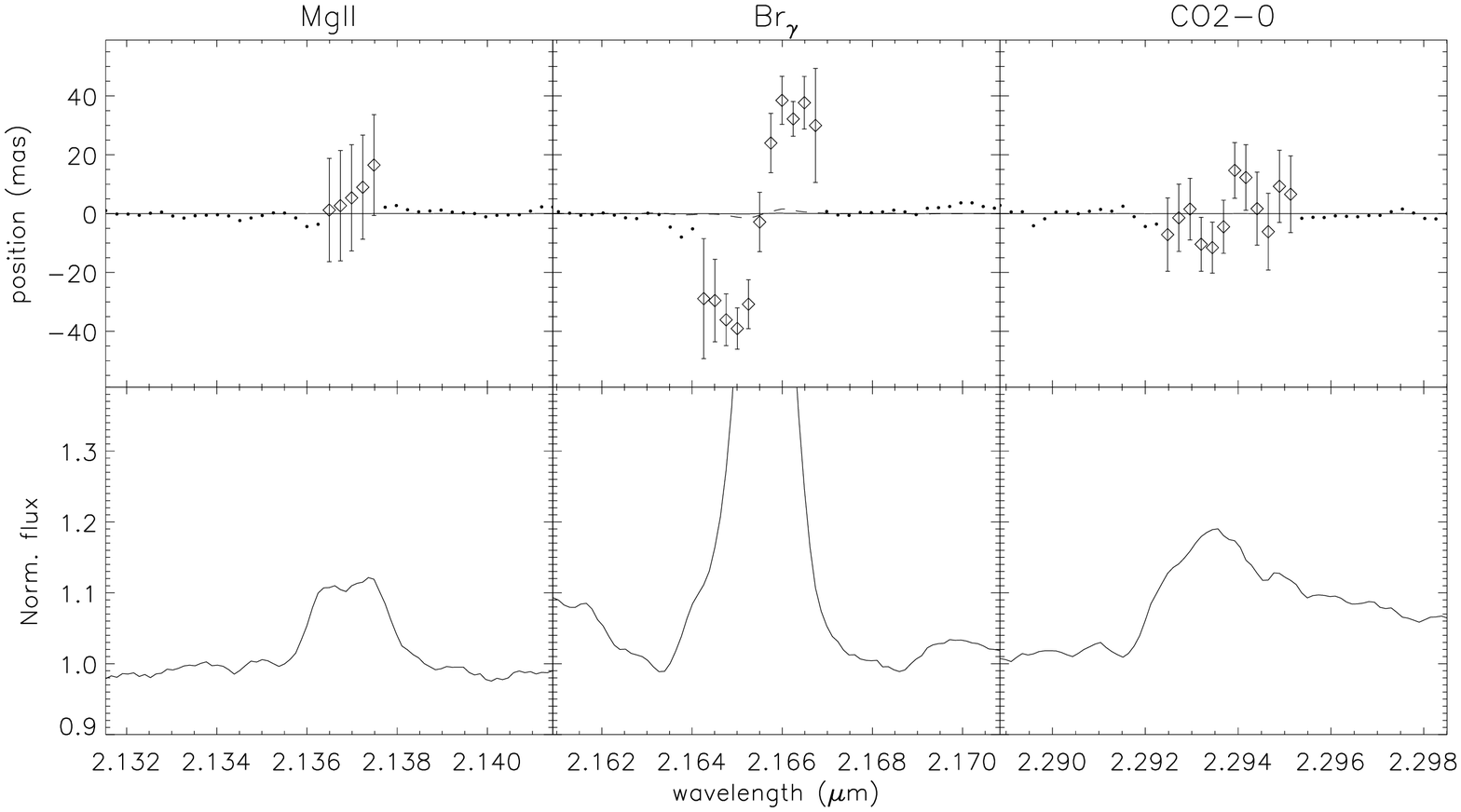}
  \caption{Same as Fig.~1 for the source 16164-5046nr3636. The dashed
  line in \brg\ panel shows the synthetic data computed using
  Equation.~6 of \citet{bran06}}
 \end{figure}

 \begin{figure} \centering \includegraphics[width=9cm]{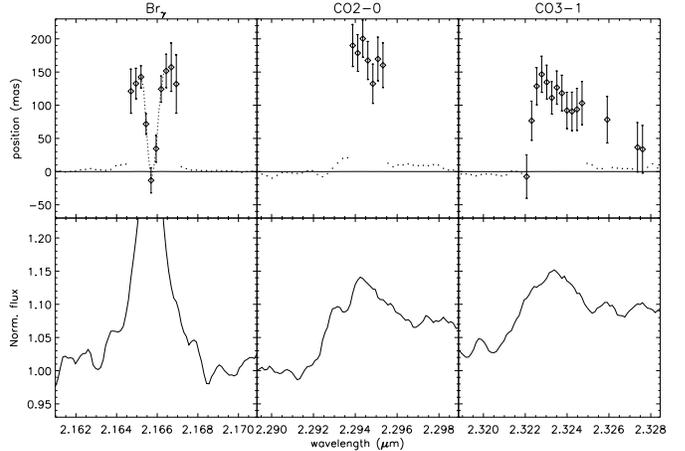}
 \caption{Same as Fig.~1 for the source 18006-2422nr766. The dotted
 line in \brg\ marks the region of residual. }
 \end{figure}

 \begin{figure}
   \centering
   \includegraphics[width=9cm]{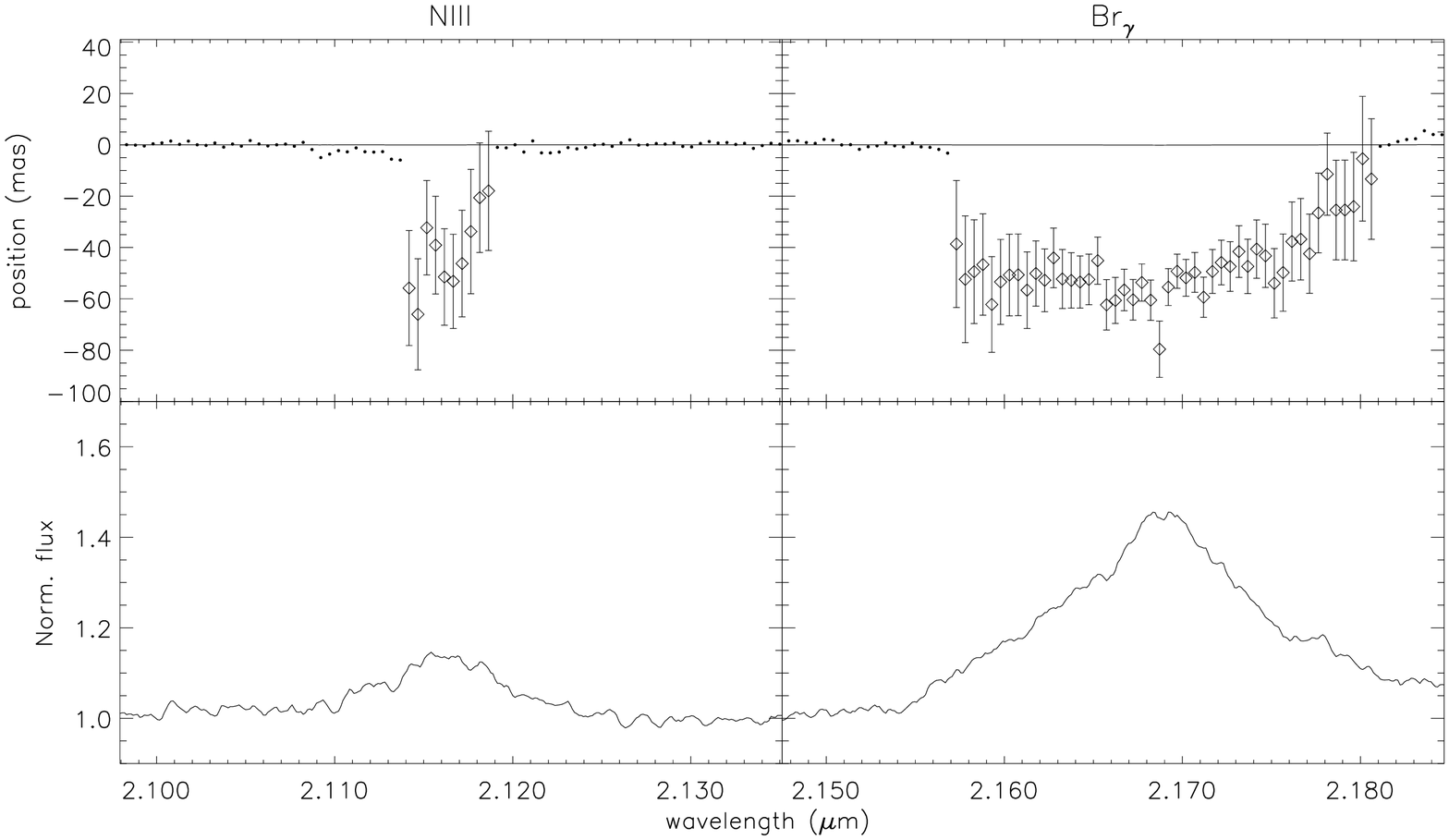}
 \caption{Same as Fig.~1 for the source 18449-0158nr335}
 \end{figure}

\section{Results}

The aim of our experiment is to constrain the physical size scales
from which different representative lines are produced in the
circumstellar environment of a young massive star. SA technique
directly measures the asymmetry of the emission line regions
about the centroidal continuum position and not the actual size. For
instance, if the geometry and kinematics of the emission line regions
are spherically symmetric, no SA signal will be detected. However, in
reality it is extremely rare to find a spherically symmetric
YSO. The environment surrounding a YSO is usually asymmetric,
due to the clumpy nature of the molecular cores from which they form
and asymmetries are more prominent around MYSOs and also commonly
observed for UCHII regions \citep[e.g.][]{kurtz04}. Even in the case of low
mass stars which are relatively more symmetric, the inclination of the
YSO to the line of sight, in combination with the extinction provided
by the circumstellar envelope will invariably produce an asymmetry in
the projected plane of viewing \citep{Whit03}. This asymmetry can separate out as a SA signal if the envelope is in emission rather than simply scattering the star light as in the case of \citet{Whit03}. It is so, because the emission line centroid will be a result of envelope geometry alone whereas the continuum centroid will be due to the star$+$envelope contributions for the observer.

Alternatively, SA displacements can also be attributed to companion emission line
stars. As we shall discuss in the next section, this is a less likely
scenario for the sample of MYSOs. CO band head emission is
known to arise in the disks of low mass stars \citep{carr89} and
recently shown to be present in MYSOs \citep{Bik06}. The number
densities in MYSOs are generally of the order of
10$^6$-10$^7$~cm$^{-3}$ \citep[e.g.][]{mue02} and can peak by a couple
of orders of magnitude locally for the luminous sources with
L$>$10$^4$\lsun, implying that the inner regions of envelopes around
MYSOs have the required densities to produce CO emission. SA displacements
measured from \brg, CO band head emission lines can therefore
constrain the physical dimensions of inner ionised regions and disks
around MYSOs. Similarly, H$_{2}$ lines which are representative of
shocked or fluorescent emission \citep{Smith94} can constrain sizes of
jets/outflows and/or accreting columns of the envelope.
 
In Figures.~1--4 the lower panel shows the intensity spectra and the
upper panel shows the position spectra for the sources
08576-4334nr292, 16164-5046nr3636, 18006-2422nr766 and 18449-0158nr335
respectively. The various columns displays data for different emission
lines as indicated on the top of each column. It can be seen that the
SA displacements have various shapes and magnitudes for different
sources and emission lines. The non-repetitive nature of the SA signal
in various sources and for different emission lines, despite all the
observations originating from the same instrument and telescope
configuration, allows us to believe that these SA signals are real
rather than possible artifacts from the applied method of
analysis. The error bars in each figure represent the errors of
Eq.~2. It can be seen from these figures that most emission lines
display SA signals. Any displacement that is larger than the measured
uncertainties of the fitting is treated as a true signal.

The \brg\ SA signal from the source 16164-5046nr3636 imitates the
shape of an artifact described by \citet{bran06}. We therefore
used the intensity spectra of this source, assumed a seeing of
0.9\arcsec and a PSF similar to be observed PSF and compatible with a
binary of maximum separation of 0.84\arcsec and generated a synthetic
SA signal using Eq.~6 of \citet{bran06}. This signal shows an angular
displacement of 0.003\arcsec and it is overplotted on the observed \brg\
signal in Fig.~2. Since the latter is much larger in magnitude than
the synthetic model of artifact, we believe that this displacement is
real.

\begin{table*} 
  \caption[]{Details of the studied sources.}
  \begin{tabular}{@{}lrrrrrrrrrrr@{}}
  \hline
      Name & R.A.(2000) & Dec.(2000) & $d$ & fwhm  & \multicolumn{4}{c}{displacement (AU)} & $M$ & $i$& $T_{ex}$ \\
        &  &  & (kpc) & ($\arcsec$)  & NIII & Br$\gamma$ & CO 2-0 & CO 3-1 & $(M_{\odot})$ & $(\circ)$ & $(K)$\\
\hline
  08576-4334nr292 & 08:59:21.58 & -43:45:31.5 & 0.7 & 0.60,0.47  & -&-& 13$\pm$4&-&6&27&1660\\
 16164-5046nr3636 & 16:20:11.31 & -50:53:25.3 & 3.6 & 0.92,0.63 &-& 279$\pm$10& 134$\pm$11&-&30&30& 4480\\
 18006-2422nr766 & 18:03:40.29 & -24:22:39.6 & 1.9 & 0.83,0.84 &-&247$\pm$23&311$\pm$32&210$\pm$39&11&60&1800\\
 18449-0158nr335 & 18:47:36.66 & -01:56:34.0 & 4.3 & 1.08 & 279$\pm$21&370$\pm$13&-&-&-&-&-\\
\hline
\end{tabular}
\vskip 1mm
The $M$, $i$ and T$_{ex}$ values are estimates from \citet{Bik04}.
\end{table*}

The SA technique measures the centroid of the line emitting regions
that are expected to trace the mid-points and/or centre-of-gravity
points, thus providing lower limits on the actual dimensions of the
projected view. This is true only when the centroidal position
of the continuum is located within the emission line region as is most
likely in the case of observed MYSOs. Although all the analysed data
are shown in the figures, only the true signals as defined above are
translated to projected physical dimensions and listed in Table.~1. In
the following we describe possible interpretations of the SA signal in
each source.
\vskip 1mm
{\bf08576-4334nr292:} This source shows H$_2$, \brg\ and CO 2-0 line
emission, representative of shocked/fluorescent gas, ionised
region and a disk/envelope, but the SA signal is not prominent according to the
criteria described above. The CO 2-0 line SA signal may be real since
the average positional displacement is slightly above the 3$\sigma$
limit. The displacement is spread on either side of the continuum
reference which measures 13$\pm$4 AU when projected to the source
distance of 0.7 kpc.
\vskip 1mm
{\bf16164-5046nr3636:} This is the most massive of the three MYSOs
examined in this work. Along with \brg\ emission, this source also
displays MgII line representative of high temperature and possibly
representing an O star candidate. Only the \brg\ line shows a
significant SA signal with the blue-shifted and the red-shifted parts
of the line profile displaced in opposite directions with respect to
the continuum source. Such a profile can be the result of an expanding
shell or a bipolar outflow. Since \brg\ lines are known to trace
ionised material, and not found in outflows, the most likely
explanation is therefore an expanding shell around the central star
due to an UCHII region or a wind.
\vskip 1mm
{\bf18006-2422nr766:} This source displays intense \brg\ and CO
bandhead emission indicative of an ionised region and a molecular
disk/envelope. The source is embedded in the midst of a busy massive star
forming HII region, namely M8. The SA signals measured from \brg\ , CO
2-0 and CO 3-1 are all significant and are found to be displaced
towards one side of the source. The projected dimensions of these
displacements are 247$\pm$23, 210$\pm$39 and 311$\pm$32 AU for \brg,
CO 3-1 and CO 2-0 respectively. These size scales suggest a scenario
where the central star is surrounded by an ionised shell, both of
which are enclosed in a molecular envelope made of CO, with the CO 3-1
emitting region placed interior to the CO 2-0 emitting
region. However, the displacement of all these emissions towards one
side of the continuum suggests that these shells/envelopes are partially
extincted or display an asymmetrical shape about the continuum. We
propose two alternate scenarios to explain this
result. Asymmetric emission due to local clumping or due to
the inclination effect described in the beginning of this section can
explain the observed SA signal for an inclination angle in between a
pole-on and edge-on configuration \citep[see Fig. 12a in][]{Whit03}. In
this particular case, although the star and the surrounding UCHII
region can themselves be bright at 2\mic, a large CO envelope with an
inner cavity can provide significant extinction at
intermediate line-of-sight angles to imitate the effects discussed by
\citet{Whit03}. Alternatively, YSOs immersed in intense UV fields
caused by HII regions are prone to produce elongated envelopes on one
side such as proplyds in the Orion Nebula \citep[see for example simulations by][]{RY00}. Interestingly, a bright OB star is situated 3.5\arcsec\
away (equivalent to a projected distance of 7000 AU) to the opposite
side of the line emission displacements, as observed on the slit. But,
this star could be a chance coincidence in the line of sight and the
available data is insufficient to conclude on any particular scenario.
\vskip 1mm
{\bf18449-0158nr335:} This source is a Wolf-Rayet star in the observed
sample displaying the emission lines CIV, NIII and \brg. No data is
available to ascertain the presence/absence of any lines
characteristic of molecular emission. The CIV line is not bright
enough to carry out SA analysis but the NIII and \brg\ lines show
reasonable SA signal projecting to size scales of $\sim$ 300 and 400
AU respectively and displaced towards one side of the
source. Wolf-Rayet stars are known to display asymmetrical mass loss
due to fast rotation and/or companion stars. There is a faint star
located 1.5\arcsec\ to the south of this source (also seen on the
slit) emitting CIV and \brg\ but not the NIII line. These emission
features common to both stars and a projected separation of 7500 AU
may suggest that the faint star is a true companion to the primary
Wolf-Rayet source. However, more concrete evidences are required to
arrive at a conclusion.

\section{Discussion}

We have tested the applicability of the technique of
spectro-astrometry to study the close environment of massive stars
using a small sample of archival VLT spectra. As shown in the previous
section, characteristic emission lines from ionised and molecular
regions display different SA displacements, constraining the physical
dimensions of the emitting regions. In a classical application of SA
technique to study spectroscopic binaries and nearby low mass YSOs the
kind of SA displacements shown in Figs.~3 \& 4 would be attributed to
the presence of companion stars with line emission. In the present
case, the analysis is that of a sample of MYSOs and emission lines
characteristic of high temperatures that can be produced in extreme
conditions involving ionization. While MYSOs are generally associated
with clustering, close companions at the level of a few hundred AUs
are not known. Companions due to fragmentation at similar size scales
close to a MYSO, which is surrounded by hot gas is also not viable
according to the Jeans criteria \citep[e.g.][]{kumar03} . At the
typical distances of a few kpc at which these MYSOs are located, only
a very luminous object such as another MYSO can produce the observed
\brg\ line intensities. Therefore, we do not attribute the SA displacements
of the MYSOs to binary companions, although it will be an interesting
experiment to pursue for future studies.

Recall that these size measurements are lower estimates on the actual
dimensions since the SA technique measures the centroid of the
emission rather than the edges. The listed values in Table.~1
($\sim$10$^2$AU) are similar to the estimated sizes of inner disks
according to numerical simulations of MYSOs \citep[e.g.][]{York02} but
smaller than the sizes of toroids ($\sim$10$^3$AU) observed around
MYSOs \citep{Cesa99,Belt05}. In comparison, the estimates from
\citet{Bik04} are much smaller (few AU). A proper understanding of the
true nature of the circumstellar environment of the MYSO can be made
only by combining the different views obtained from multiple methods
of investigation from the millimetre to the NIR wavelengths.  The SA
measurements can be very useful to provide the size scales based on
NIR observations tracing the very hot and inner regions of the MYSOs
and provide alternative ways to evaluate the consistency of results
obtained using other methods. For example, the three MYSOs examined in
this work have been analysed by \citet{Bik04}. These authors modelled
the CO line profile of the band head emission in the framework of
Keplerian rotation. They arrive at a disk radii of $\sim$3--4 AU for
the sources 16164-5046nr3636 and 18006-2422nr766, for certain assumed
inclination angles. In contrast, our SA analysis for the same sources
indicates radii of a few hundred AU which are representative of much larger disks (not necessarily Keplerian) and/or envelopes
around MYSOs. Also, the SA method yields different envelope radii
traced by CO 2-0 and CO 3-1 band heads for the source 18006-2422nr766,
implying the origin of these lines from regions with different
physical conditions. These contrasting results between SA analysis and
line profile modelling may imply a scenario where the massive young
star is surrounded by an optically thick inner disk and an outer
envelope.

The experimental application of the SA method here demonstrates the
potential of this technique in revealing the details of the close
environment of MYSOs. The results presented here are only indicative,
based on spectra obtained in one position angle. If systematic SA
observations in multiple position angles are obtained on a larger
sample of MYSOs, the results can prove to be of invaluable importance
in understanding the complex environment of MYSOs. The SA technique is
also an efficient and inexpensive method that needs only limited
amounts of telescope time. In contrast techniques such as infrared
interferometry with very large telescopes and/or mm wave
interferometric imaging consume large amounts of telescope time and
analysis efforts to probe the same regions.

\begin{acknowledgements}
We gratefully acknowledge Paulo Garcia for a code that served to
verify against artifacts. We also thank the referee M. Takami for
useful comments that clarified several aspects of the paper. JMCG and
MSNK are supported by a research grant POCTI/CFE-AST/55691/2004 and
JMCG is supported by a doctoral fellowship SFRH/BD/21624/2005 approved
by FCT and POCTI, with funds from the European community programme
FEDER. This work is based on data obtained with ESO facilities at
Paranal that was retrieved using ESO Science Archive Facility.

\end{acknowledgements}

\end{document}